\documentclass[aps,epsf,preprint,showpacs]{revtex4}
\usepackage{graphics}
\usepackage{graphicx}
\begin{document}
\newcommand{\dr}{\delta \rho}
\newcommand{\h}{\cal {H}}
\newcommand{\U}{\cal {U}}
\newcommand{\beq}{\begin{equation}}
\newcommand{\eeq}{\end{equation}}
\newcommand{\bfig}{\begin{figure}}
\newcommand{\efig}{\end{figure}}

\title{Nonlinear charging, and transport times in doped nanotubes junctions}

\author{Keivan Esfarjani}
\affiliation{Dept. of Physics, Sharif University of Technology,
11365-9161 Tehran  IRAN}
\email{k1@sharif.edu}

\author{Amir A. Farajian}
\affiliation{IMR, Tohoku University, Sendai 980-8577 JAPAN}
\author{Siu Tat Chui}
\affiliation{BRI, University of Delaware, Newark, DE 19716 USA}
\author{Yoshiyuki Kawazoe}
\affiliation{IMR, Tohoku University, Sendai 980-8577 JAPAN}

\date{\today}

\begin{abstract}
The nonlinear capacitance in doped nanotube junctions is
calculated self consistently. A negative differential capacitance
is observed when the applied bias becomes larger than the
pseudogap of the metallic armchair nanotube. For this device, one
can deduce a relaxation time of about 0.1 femtosecond. Because of
its negative differential resistance (NDR), a switching time of
less than a fs, i.e. at least 3 orders smaller than present day
switching times, can also be estimated. This effect is important
in designing ultra fast nano-electronic components.


\end{abstract}
\maketitle

Proposals for the calculation of the quantum capacitance of a
mesoscopic system were put forth by B\"uttiker\cite{buttiker} and
coworkers and later its full quantum version was derived by Zhao
et al. \cite{zhaoguo} where the correction terms to the
geometrical capacitance, unlike the result derived previously by
B\"uttiker, could become negative due to effects such as quantum
tunneling (a ``leaky" capacitor). The work, mainly by B\"uttiker
and coworkers\cite{buttiker}, on mesoscopic systems has studied
the RC relaxation time in the presence of an applied AC field.
They had already noted that the true capacitance in a mesoscopic
sample has contributions from two hypothetical capacitances put in
series: the first being the geometric one caused by the Coulomb
interaction which also persists in macroscopic samples. The second
is a material-dependent effect caused by the filling of the
(quantized) states of the system; it exists even for non
interacting electrons, and becomes negligible for large metallic
systems where the density of states (DOS) at the Fermi level
becomes large.

Experiments observing a negative value for the capacitance of a
metal-carbon nanotube junction in the AC regime have been
performed by Zhao et al. \cite{zhao2}. A non classical capacitance
was also measured by Hou et al. \cite{hou} in single electron
spectroscopy of a double barrier tunnel junction. In their setup,
for a small tip-cluster separation, the non classical behavior
(reduction) of the capacitance and the resistance was interpreted
in terms of quantum tunneling of electrons.
Theoretical works on the ac-response of one-dimensional systems
have also been done. These models\cite{sassetti} use the Luttinger
liquid theory, which is the effective low-energy theory of one-dimensional
systems, to describe screening and ac response in homogeneous 
1D quantum wires. It is not clear, however, whether at high biases
such theories remain valid since the linear dispersion at the two
ends of the junction is not the same anymore, and furthermore, in this 
paper, we are considering junctions of {\bf doped} nanotubes. 

In this work, we have a fully quantum mechanical theory explaining
the decrease in C which is closely related to the electronic band
structure of the tube. Also, from our calculations, the relaxation
time of the nanotube junction  is found to be extremely fast, of
the order of a fraction of a femtosecond. Additionally, the
switching time of this device is estimated to be of the same order
as well. These features are central for the design of nanoscale
devices, as the switching time issue is of paramount importance in
making fast chips. 

In the presence of a time-dependent applied bias, the quantum
capacitance and transport times in doped junctions made of carbon
nanotubes have been calculated in a self-consistent manner. In our
model, this junction can be made by a nanotube deposited on or
embedded in two different materials. Depending on the material,
this can induce transfer of electrons to or from the tube.  We had
previously studied the I-V characteristics of a nanotube junction
in the DC case\cite{apl,prl} and found that the current is a
non-linear function of the applied voltage. For semiconducting
tubes, a rectifying behavior was observed, whereas metallic tubes
displayed a negative differential resistance (NDR) specially for
narrow tubes where the quantization of the transverse levels
(pseudogap) is more significant. For an applied AC voltage, in
addition to its resistance, the tube will also manifest its
capacitive properties. Indeed there is a region of charge
accumulation at the junction where, due to the difference in the
chemical potentials on opposite sides of the junction, there will
be diffusion of electrons from the donor into the acceptor region,
and diffusion of holes into the donor region. It was previously
shown\cite{screening} that the depletion-layer length of such a
junction is of the order of a few carbon-ring separations. The
doped nanotube junction can thus be viewed as a capacitor with a
capacitance C in parallel with a resistance R if one assumes that
the potential drop occurs in the junction region. The response
time of this junction is usually controlled by the time $\tau_{\rm
relax} = RC$ where R is the resistance of the junction. Hidden
beneath the apparent simplicity of this expression is the quantum
electron transport time across the junction. Our exact quantum
calculation of R\cite{prl} has incorporated this effect. In this
work, we focus on the calculation of C and show that it is a
nonlinear function of the voltage whose differential value can
become negative. Furthermore, from the I-V and the Q-V
characteristics, a switching time is defined and estimated for
this junction. In what follows, the model and the calculation
method will be described.

We consider an infinite ideal (defectless) nanotube doped with two
different dopants on its left and right. The doping can be
realized by either inserting dopant atoms inside the tube, by
depositing the tube on a substrate, or by embedding it in a host
material. In both cases, there will be a charge transfer to or
from the tube which will now acquire different electronic
properties due to the addition of charge and the eventual shift in
its Fermi level. It is assumed that there are no charged
impurities present in the proximity of the junction, and therefore
the effect of disorder is neglected in this study.

The junction problem is solved in the real space within a one
$\pi$ orbital tight binding (TB) formalism. First the two isolated
half tubes are treated, i.e. their surface Green's function
($G_{00}$) in the tight-binding basis is calculated by using the
renormalization method\cite{lopez-sancho}, assuming the effective
onsite energies and chemical potentials known\cite{explain}. Then
the junction problem is solved by using the Green's Function (GF)
matching method\cite{garcia}. This is done self-consistently in
the junction region, by adding to the Hamiltonian, a Coulomb
interaction term (Hartree potential) due to both onsite and other
sites extra charges. The onsite (Hubbard) term $U \approx
11$eV\cite{scf} is about 4 times the hopping integral $t$. We must
emphasize that the calculation of the transmission coefficient
from which the two-terminal resistance of the junction is derived
is an exact calculation within the self-consistent TB model.

A (3,3) and a (5,5) armchair nanotube\cite{iijima} were considered
in this calculation as the nonlinear effects are more pronounced
in small radius tubes. The junction region includes 8 unit cells
and is of length $19.4 \AA$. The onsite energies on the left and
right tubes are respectively $U_L$ and $U_R$. The unit for energy
or voltage is the hopping matrix element $t=2.7 eV$.

From the GF, one can compute the conductance, and the charge
distribution in the junction region as a function of the applied
bias. Thus both relations $Q(V)$ and $I(V)={\dot {Q}}(V)$ can be
obtained from this calculation. Fig. \ref{charge} shows the
depletion charge profile of a (5,5) tube in the junction region
which contains 8 layers (16 carbon rings). It was checked that
increasing the number of layers did not change the amount of
charge transfer on these layers\cite{screening}. The depleted
charge is calculated for two different dopings of $|U_L|=|U_R|=0.1
t$ and $0.5 t$.

%
%

For large voltages, the charge-bias relation becomes nonlinear.
The bandwidth of the local density of states (LDOS) in the
junction region is larger than that of the bulk tube by the amount
of the bias voltage. As a result, after the bias becomes greater
than the pseudogap of the tube, the accumulated charge is reduced.

This nonlinear relation is displayed in Fig. \ref{ivqv} where t he
average charge is the sum of the charges on all rings on one side
of the junction. For large biases, the charge on each side of the
junction is not strictly zero but oscillates along the tube axis
so that its average value becomes nearly zero. This oscillation
plays an important role in determining the characteristics of the
device since it causes the charge on one side of the capacitor to
become small and yield a small switching time. The nonlinear
$I(V)$ curve is also displayed in this figure.

%
%

We calculate the linear capacitance as the ratio of the charge
induced on the two sides of the junction, divided by the applied
voltage: $C=Q/V$. For small bias voltages, the charge-bias
relation is linear, and, assuming a hopping of 2.7 eV and the
slope to be 0.55 electron/hopping, we obtain a capacitance of $2.8
\,10^{-20}F$. This is comparable but smaller than the
``geometrical'' capacitance\cite{comment} 
$C_g=\epsilon A/d \approx 3.6
\times 10^{-20}F$, where $A$ is the area of the nanotube (ring of
circumference 21 \AA\ and thickness 3.3\AA ), $d$ is the distance
between the dipole layers (about 5\AA ) and $\epsilon=3$ is the
in-plane dielectric constant of graphite\cite{dresselhaus}.

Taking the resistance $R$ to be $6.5 k\Omega$\cite{comment2}, we
find a {\bf very fast} relaxation time of $\tau\approx 0.2$ femto
second. The resistance of a nanowire is always of the order of
$h/2e^2=13 k\Omega$, and the capacitance is about $9 \times
10^{-12} A/d \approx 10^{-11} 10^{-9} = 10^{-20} F$, resulting
always in a relaxation time of the order of $10^{-16} s$, or,
including the dielectric constant, a fraction of a femto second. A
smaller separation and sources of disorder may increase this value
by an order of magnitude.

The fast relaxation time in junctions of nanowires comes about not
only because the capacitance is very low but also because the
resistivity is so small. Carbon nanotubes have a very small
resistivity $\rho$, and furthermore, their resistance does not
increase with length. We emphasize the resistivity of the junction
is related but not the same as that of the carbon nanotube. It
comes from a detailed sophisticated quantum transport calculation.
That the magnitude is so small has not been perhaps emphasized
enough previously. In general the resistance is inversely
proportional to the area, $R\propto \rho/A$, whereas the
capacitance is proportional to $A$. The time constant $RC$ is thus
{\bf generally independent} of $A$ no matter how small the device
is! However, in the case of nanotubes only, the conductance of the
tube is of the order of $2 \times 2e^2/h$ and thus independent of
its cross sectional area.

From the knowledge of the relationship between the charge $Q(V)$
and the current ${\dot {Q}}(V)$ as a function of the bias, one can
obtain the dynamics of this device. The slope at zero (small bias)
is the relaxation time $\tau=RC$ already discussed. The other
characteristic time which can be observed at biases where the NDR
effect is seen is called the switching time. It is the time
necessary for switching from "zero" current (in the I-V valley) to
its maximum value.

The switching time can be observed as the other characteristic
time scale in the plot of charge versus current after the current
has reached its maximum value. In Fig. \ref{phase}, it can be
identified as the large, almost vertical slope seen on the right
side of the s-shaped curve. According to our calculations, it is
less than an order larger than the relaxation time ($\tau_{\rm
switching} \approx 4.5 \tau_{relaxation}$ for the (3,3) tube). It
is therefore still of the order of a femtosecond.


In general, the switching speed is enhanced if the peak to valley
ratio of the NDR characteristic is large. Indeed this would
stretch the curve in the ${\dot Q}$ direction, and therefore
increase the denominator $\Delta {\dot Q}$, thereby reducing the
switching time. As the device becomes comparable in size to the
mean free path or larger, the resistance starts increasing with
length, and both the relaxation time and the switching time
increase. So for nanoscale systems where there is quantum
coherence, one expects to obtain very fast devices capable of
handling high switching speeds.

Inductive effects in this device have been neglected. The
inductance $L$ for a device of size $l$ is of the order of $\mu_0 l$
and thus the impedance is about $L \omega \approx \mu_0 l \omega \approx
10 \Omega$ for a nanometer-length device at frequencies of 10$^{16}$ Hz.
This number is still a thousandth of the resistance of the junction,
and therefore can be justifiably neglected.

There are three effects that can affect the relaxation: Impurity
scattering, electron-phonon scattering and radiation. Carbon
nanotubes have no impurities in their structure unless made on
purpose. The electron mean free path being of the order of 1000 nm
or more in nanotube ropes, it is very unlikely to find impurities
within or near the junction which is of a much smaller size. Their
potential is therefore very smooth and will not affect the
properties of the electrons in the junction region.

Electron-phonon interactions will increase the resistance, but
this increase is by a factor of the order of unity (see ref.
\cite{e-phonon} for example). The other  effect which becomes
important at such high frequencies is radiation. The radiated
power is proportional to the second power of the acceleration of
electrons which is itself quadratic in frequency. At high
frequencies, one thus expects some power loss by radiation. The
radiated power by a dipole is approximately $P= v^2 \omega^2 \beta
$ with $\beta \approx \sqrt {\mu_o/\epsilon_o} \,e^2/c^2 $. We
assume that the shape or geometry of our device affects this
coefficient by a factor of the order of one. The radiated power
being quadratic in the velocity, one can associate a damping force
to it: $F_{\rm damping}=P_{\rm rad} /v = - m v/\tau'$ with a
relaxation time $\tau' = m/\beta \omega^2 = \sqrt {\epsilon_o/
\mu_o} \, mc^2/e^2 \omega^2$. This time should be compared to the
relaxation time at frequencies of the order of 10$^{16}$ Hz. A
numerical substitution shows that $1/\tau' \approx 10^9 {\rm Hz}
\ll 1/\tau \approx 10^{16}$ Hz. Thus, the relaxation time due to
radiation does not depend much on the characteristics of the
system except for the $\omega^2$ dependence, and radiation does
not affect the relaxation time either.

In summary, we find that one can define two different time scales
for such a junction displaying NDR and nonlinear charging
properties: 1) the relaxation time (RC) which is just the time
necessary for the initially charged junction to discharge provided
that the bias is small enough so that the device characteristics
are linear; 2) the switching time which is the time needed for the
voltage to transit from the NDR valley to the linear region. For
the doped nanotube junction considered here, both times are very
short because the resistivity of carbon nanotube is very low. This
makes it attractive for possible practical applications.

\begin{acknowledgments}
We wish to acknowledge the supercomputer center of IMR for their
continuous support of the HITAC supercomputing system.
\end{acknowledgments}

\newpage

\bfig
\includegraphics[width=14cm]{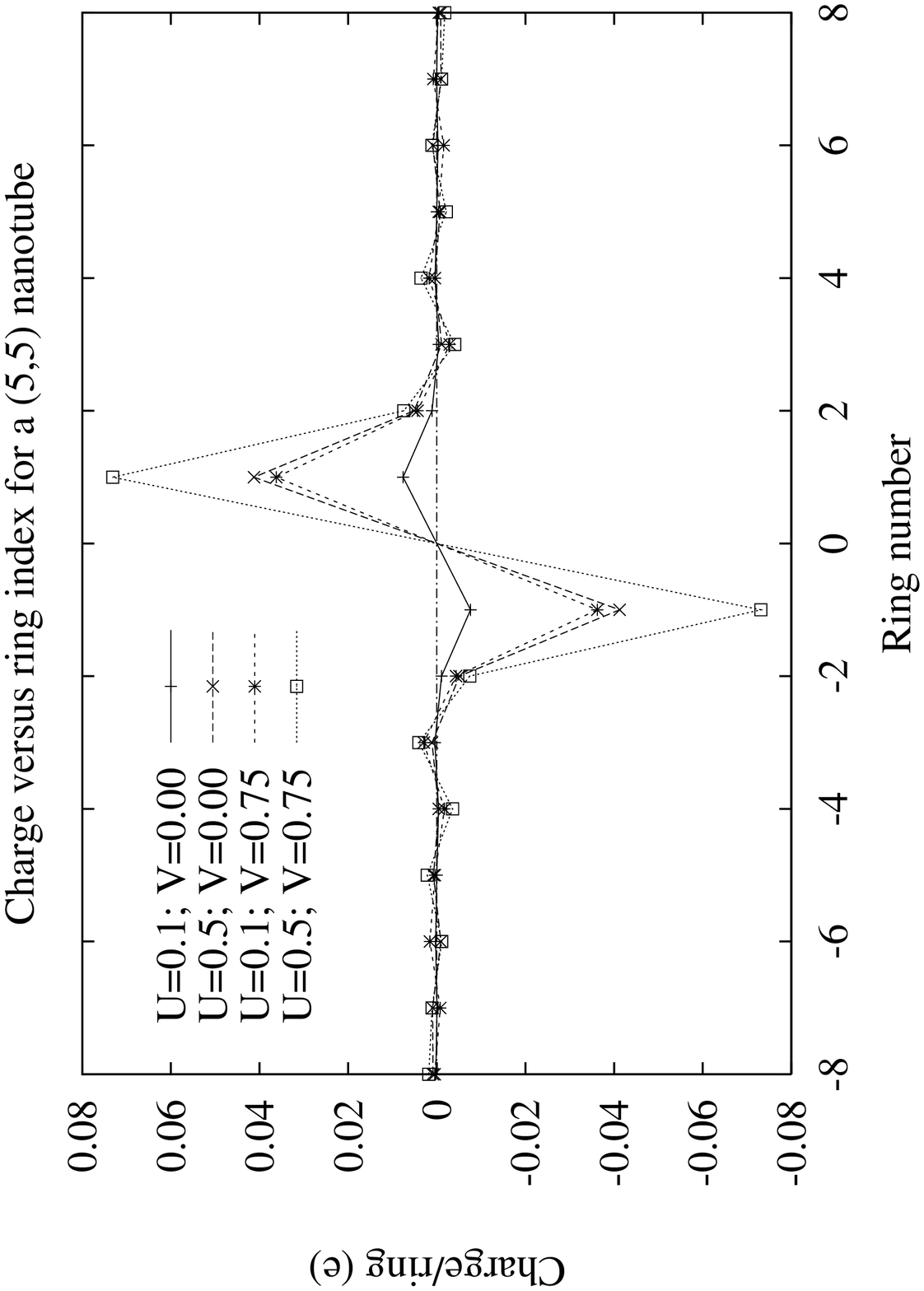}
\caption{Charge transfer across the junction in a (5,5) armchair
tube under the bias of 0 and 0.75 hopping, for two different
dopings of $U_{l,R}=pm 0.5$ and $U_{l,R}=pm 0.1$. The central cell
contains a total of 8 layers or 160 atoms} \label{charge}
\efig 

\newpage
\bfig
\includegraphics[width=14cm]{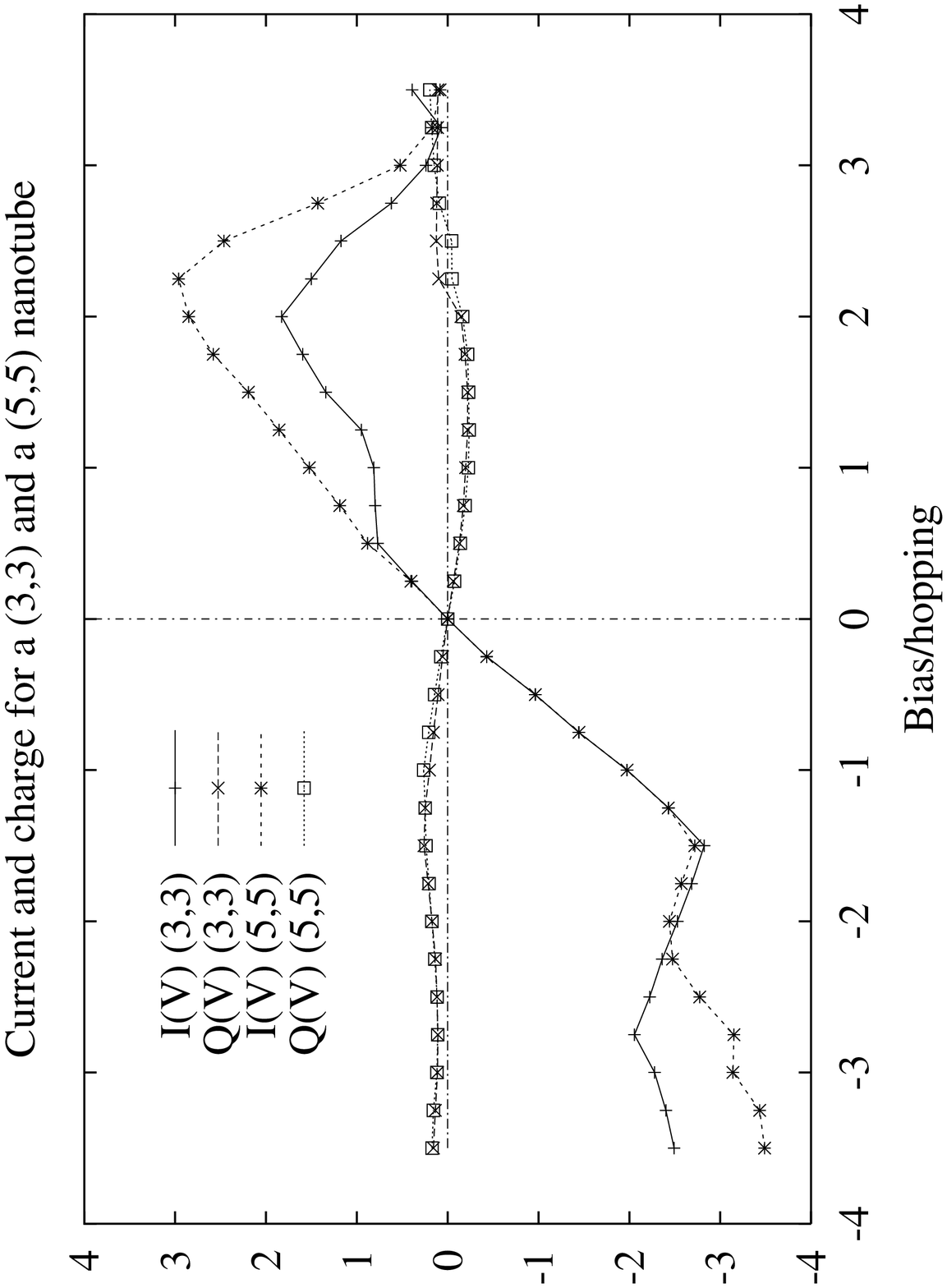}
\caption{$Q(V)$: Average charge (e) and I(V): current in $2 e/h
\times t$ versus the applied bias (in $t$) for a (3,3) and a (5,5)
nanotube with onsite energies of $-0.5 t$ for the left and $0.5 t$
for the right side tube. The residual dipole charge at zero bias
at the interface due to doping has been subtracted in order to
have curves passing through the origin.} \label{ivqv}
\efig 

\newpage

\bfig
\includegraphics[width=14cm]{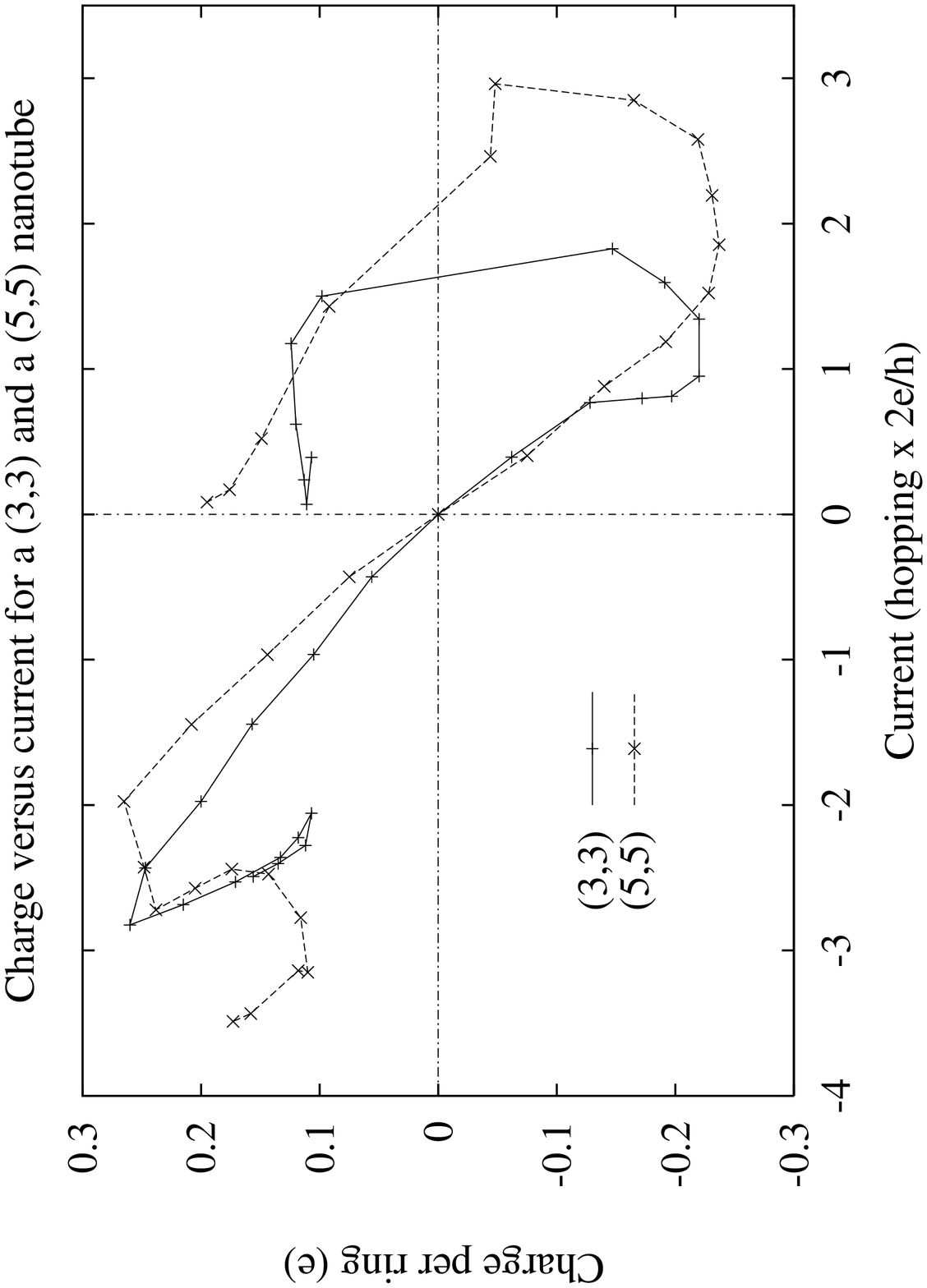}
\caption{Phase space trajectory ($Q,{\dot{Q}}$) for both (3,3) and
(5,5) tubes at $pm 0.5 t$ doping level. It is asymmetric due to
the asymmetry caused by N-P doping } \label{phase}
\efig 


\begin{thebibliography}{99}
\bibitem{explain}
The effect of the substrate or dopants is to transfer some charge
to the tube. This modifies the onsite energies which we assume to
be known for the two semi-infinite tubes. From this, the
additional charge can be easily deduced without any need to
perform a self-consistent calculation, if the onsite energies are
uniform. At the junction, however, a self-consistent calculation
needs to be performed in order to determine the charge and
potential profiles, even if the two separated half tubes are
treated non-selfconsistently.
\bibitem{comment}
$C_{\rm equivalent}^{-1}=C_{\rm geometrical}^{-1}+C_{\rm quantum}^{-1}=
\epsilon A/d+e^2 DOS(\mu)$. 
This relation, already explained in the introduction, explains 
why the real value of the capacitance of a nanodevice is always 
smaller than its geometrical value.
\bibitem{comment2}
The exact value of the junction resistance turns out to be very
near its ballistic value due to small charge accumulation accross
the junction and neglect of any disorder. The RC relaxation time
is obtained from this exact calculation.
\end{thebibliography}
\end{document}